\documentclass[useAMS,usenatbib]{mn2e}
\usepackage{graphicx}
\usepackage{amsmath}
\usepackage{amssymb}
\newcommand  \acc     {\ifmmode {\rm km\,s}^{-2} \else km\,s$^{-2}$\fi}

\newcommand  \ergs     {\ifmmode {\rm ergs\,s}^{-1} \else ergs s$^{-1}$\fi}
\newcommand  \ergcms   {\ifmmode {\rm erg~cm}^{-2}\,{\rm s}^{-1}
                        \else erg~cm$^{-2}$\,s$^{-1}$\fi}
\newcommand  \ergcmsA  {\ifmmode{\rm erg\,cm}^{-2}\,{\rm s}^{-1}\,{\rm\AA}^{-1}
                        \else ergs\,cm$^{-2}$\,s$^{-1}$\,\AA$^{-1}$\fi}
\newcommand  \ergcmsHz {\ifmmode{\rm ergs\,cm}^{-2}\,{\rm s}^{-1}\,{\rm Hz}^{-1}
                        \else ergs\,cm$^{-2}$\,s$^{-1}$\,Hz$^{-1}$\fi}
\newcommand  \phcms    {\ifmmode {\rm ph\,cm}^{-2}\,{\rm s}^{-1}
                        \else ph\,cm$^{-2}$\,s$^{-1}$\fi}
\newcommand  \phcmsA   {\ifmmode {\rm ph\,cm}^{-2}\,{\rm s}^{-1}\,{\rm\AA}^{-1}
                        \else ph\,cm$^{-2}$\,s$^{-1}$\,\AA$^{-1}$\fi}
\newcommand\aj{{AJ}}% 
\newcommand\araa{{ARA\&A}}% 
\newcommand\apj{{ApJ}}% 
\newcommand\apjl{{ApJ}}% 
\newcommand\apjs{{ApJS}}% 
\newcommand\aap{{A\&A}}% 
\newcommand\aaps{{A\&AS}}% 
\newcommand\mnras{{MNRAS}}% 
\newcommand\pasp{{PASP}}% 
\newcommand\nat{{Nature}}% 
\title{
On the fraction of intermediate-mass close binaries that explode as
          type-Ia supernovae}
\author[Dan Maoz]
{Dan~Maoz$^{1,2}$\thanks{E-mail: maoz@wise.tau.ac.il 
}\\
$^{1}$School of Physics and Astronomy, 
Tel-Aviv University, Tel-Aviv 69978,
Israel\\
$^{2}$Kavli Institute for Theoretical Physics,
University of California, Santa Barbara, CA 93106-4030, USA}
\date{\today}

\begin{document}

\maketitle

\label{firstpage}

\begin{abstract}
Type-Ia supernova (SN-Ia) explosions are thought to result from a
thermonuclear runaway in carbon-oxygen white dwarfs (WDs) that
approach the Chandrasekhar limit, either through accretion from a
companion or a merger with another WD. However, it is unknown 
which of these channels operates in reality, and what 
are the details of the process.
I compile from the literature 
observational estimates of the fraction $\eta$ of
intermediate-mass stars that eventually explode as SNe-Ia,
supplement them with several new estimates, and compare them
self-consistently.
The estimates are
based on five different methods, each  
utilising some observable related to the SN-Ia rate,
combined with assumptions regarding the initial mass
function (IMF): 
the ratio of SN-Ia to core-collapse explosions in star-forming galaxies;
the SN-Ia rate per unit star-formation rate; the SN-Ia rate
per unit stellar mass; the iron to stellar mass ratio 
in galaxy clusters; and the 
abundance ratios in galaxy clusters. The five 
methods indicate that a fraction in 
the range $\eta\approx 2-40\%$
of all stars with initial masses of $3-8 M_\odot$ (the progenitors of the WDs
generally thought capable of exploding through the above scenarios)
explode as SNe-Ia. A fraction of $\eta\approx 15\%$
is consistent with all five methods for a range of plausible IMFs. 
Considering also the binarity fraction among such stars,
the fraction of binaries with the appropriate mass ratios,
the fraction in close initial orbits, and duplicity
(every binary can produce only one SN-Ia explosion), this
implies that nearly every intermediate-mass close binary ends up as a 
SN-Ia, or possibly more SNe-Ia than progenitor systems.
 Theoretically expected fractions are generally one to two orders 
of magnitude lower. The problem could be solved: if all the
observational estimates are in error; or with a 
``middle-heavy'' IMF; or by some mechanism that strongly enhances
the efficiency of binary evolution toward SN-Ia explosion; 
or by a non-binary origin for SNe-Ia. 
 
\end{abstract}

\begin{keywords}
supernovae: general -- binaries: close -- white dwarfs
\end{keywords}

\section{Introduction}

Type-Ia  supernovae (SNe-Ia) play a central role in astrophysics,
not only as distance indicators for cosmography, but also as major
contributors to cosmic metal production and distribution.
The occurrence of some SN-Ia explosions in early-type galaxies,
having no trace of a young stellar population, implies that at least 
some of these events are derived from old stars, likely low-mass stars
or white dwarfs (WDs). Detailed analysis of the optical spectra 
and light curves of SNe-Ia (e.g., Mazzali et al. 2007), 
and of some nearby SN remnants 
thought to have been SNe-Ia (e.g., Badenes et al. 2006), 
are consistent with a scenario in which 
a C+O WD approaches or passes the Chandrasekhar mass limit, $M_{\rm ch}$,
initiating carbon ignition under degenerate conditions.
This leads to thermonuclear runaway and incineration of much of the WD into
iron-peak elements, among which the radioactive ones power the 
optical light curve (see, e.g., Hillebrandt \& Niemeyer
2000, for a review). 

However, the nature of the process behind 
the growth toward the Chandrasekhar mass is not known. The two 
leading competing scenarios are accretion from a close binary companion
 -- the single-degenerate (SD)
scenario (Whelan \& Iben 1973; Nomoto 1982) 
or merger with another WD, following loss of orbital
energy and angular momentum by emission of gravitational waves -- 
the double-degenerate (DD) scenario (Iben \& Tutukov 1984; Webbink
1984). Both scenarios envision passage
of the binary through at least one common-envelope phase. The complex physics
of this stage are a major obstacle in calculations of the binary
evolution of SN-Ia progenitors. In the SD case, it is also unknown what 
is the nature of the companion -- a main sequence, giant, or sub-giant
star, and what is the form of the accretion flow -- 
Roche lobe overflow or a wind. Once conditions for runaway have
been reached (and these could be sub-Chandra, Chandra, or
super-Chandra), there are various possibilities for 
where in the WD ignition occurs, and how the burning front propagates
 -- various combinations of subsonic (``deflagration'') and supersonic  
(``detonation'') propagation have been considered.
A number of authors have computed, for the various possible progenitor
scenarios, 
the ``delay time distributions'' for the time between the 
formation of a stellar population and 
the explosion of some of its members as SNe-Ia (e.g., Greggio \&
Renzini 1983; Jorgensen et al. 1997;  
Sadat et al. 1998; Madau et al. 1998; Ruiz-Lapuente \& Canal 1998; 
Yungelson \& Livio 2000; Matteucci \& Recchi 2001; 
Han \& Podsiadlowski 2004; Greggio 2005; 
Belczynski et al. 2005).
The actual co-existence of several different channels that lead to
SNe-Ia is possible. In fact, evidence has surfaced recently for two
different types of SNe-Ia,  in young and in old stellar  
populations (see \S\ref{localratio}, below).

A better, observationally guided,
understanding of what objects explode in SNe-Ia, and how it happens,
is clearly desirable for the confident use of these objects as
standard candles, and for a coherent picture of metal enrichment.
In this paper, I revisit the exploding fraction, 
a simple parameter that can be derived
from observations, and which can place constraints on the SN-Ia 
progenitor population.
 
\section{Five estimates of the exploding fraction}
Any measurement implying a
 total SN-Ia number from some stellar population, relative
to some parameter that permits normalizing the initial mass function 
(IMF) of that population, combined
with an assumption about the functional form of the IMF, can
give a direct estimate of the fraction $\eta$ of stars with initial 
masses in the range $M_{\rm min}-M_{\rm max}$ 
that eventually explode as SNe-Ia.
I compile here five types of such measurements, and the explosion fractions
that they imply, which are also summarised in Table~1.

In the DD scenario, a minimum zero-age main-sequence (ZAMS) 
mass above $M_{\rm min}\approx 3M_{\odot}$ 
is probably required for each of the members of a SN-Ia progenitor binary
system if, after the mass loss involved in post-main-sequence
 and binary evolution,
a total WD mass of at least $M_{\rm ch}$ is to be attained (e.g.,
 Branch et al. 1995; Weidemann 2000; Tutukov \& Yungelson 2002).
The maximum initial mass of the primary in the binary, $M_{\rm
  max}$, equals the
minimum initial mass for eventual core-collapse, generally thought to 
be in the range $(8-10) M_{\odot}$ (Timmes et al. 1996).
It is possible that binaries with one or both masses somewhat outside
the $(3-10) M_{\odot}$ 
range can also produce SN-Ia explosions. For example, in the 
model calculations by Tutukov \& Yungelson (2002), the WDs descended
from binaries with primaries of ZAMS mass $>5M_{\odot}$ and
secondaries of ZAMS mass of only $2.5 M_{\odot}$ can still 
sometimes merge and surpass the 
Chandrasekhar limit. Tutukov \& Yungelson (2002) also assume a
relatively high minimum mass for core collapse, $11.8 M_{\odot}$,
and therefore quite massive primaries also contribute to their calculated
SN-Ia rate. The observationally identified progenitors of several 
core-collapse SNe have best-fit masses of $(8-10) M_{\odot}$, but the
model uncertainties permit
masses as high as $(12-13) M_{\odot}$ (Maund \& Smartt 2005; Li et
al. 2006;
Hendry et al. 2006). 
Nevertheless, among all the DD mergers with total mass $>M_{\rm ch}$
in the Tutukov \& Yungelson (2002) simulations, 85\% originate from 
binaries in which both members had ZAMS masses between $(3-8) M_{\odot}$
(L. Yungelson, private communication),
and therefore this is a useful range to consider for the progenitors
of the large majority of SNe-Ia. 

I will henceforth use the following notations. Stellar masses
will be expressed in dimensionless form, as $m=M/M_{\odot}$. 
The IMF, $dN/dm$, is given in terms of stars per unit mass interval.
I denote with $m_{\rm low}$ and $m_{\rm hi}$ the low-mass and
high-mass cutoffs, respectively, of the IMF. 
I will begin by assuming the following 
``standard'' parameters: $m_{\rm min}=3$; $m_{\rm max}=8$;
a Salpeter (1955) IMF, with 
$(dN/dm)\propto m^{-2.35}$;  $m_{\rm low}=0.1$; and $m_{\rm hi}=100$.
In \S\ref{imfvary}, 
I will examine the consequences of varying these parameters.
The treatment will first focus on the DD scenario, but in
\S\ref{discussion} I will investigate how the results are affected
in the SD case. 

\begin{table}
%\scriptsize
\begin{minipage}{\textwidth}
\begin{tabular}{c|c|l}
\hline
\hline
{Method} &
{$\eta (\%)$} &
{Reference} \\
(1)&(2)&(3)\\
\hline
Ia/CC&$8-15$& Mannucci et al. (2005)\\
\hline
$B$  &$5-7$ & Dahlen et al. (2004)\\
     &$8-10$& Barris \& Tonry (2006)\\
     &$6-18$& Scannapieco \& Bildsten (2005)\\
     &$3-9$ & Scannapieco \& Bildsten (2005)\\
     &$1-1.5$& Sullivan et al. (2006)\\
     &$3.8-4.3$   & Mannucci et al. (2006)\\
\hline
$A$  &$0.8-1.7$&Mannucci et al. (2005)\\
     &$1-1.5$& Sullivan et al. (2006)\\
     &$2-6$& Sharon et al. (2007)\\
     &$2-3.5$& Mannucci et al. (2007)\\
\hline
Fe&$11-20$& Lin et al. (2003)\\ 
\hline
Abund&$14-40$& De Plaa et al. (2007)\\
\hline
\end{tabular}
\end{minipage}

\caption{Observational estimates of $\eta$, the fraction of stars with
ZAMS mass $m=3-8$ that explode as SNe-Ia. 
Column header explanations: 
(1) - Ia/CC -- ratio of the type-Ia and core-collapse SN rates
in star-forming galaxies; $B$ -- the SN-Ia rate per unit
star-formation rate; $A$ -- the SN-Ia rate per unit mass; Fe -- the
ratio of iron mass to stellar mass in galaxy clusters; Abund -- the 
number ratio of type-Ia and core-collapse SNe that have exploded in
clusters, based on abundance patterns;    
(2) - Estimate of $\eta$, in percent, after 
converting to the adopted IMF parameters (see text).
(3) - Reference for the original estimate, or the
observational data used in the prsent estimate (see text).  
}
\label{table1}
\end{table}

\subsection{The local ratio of type-Ia to core-collapse SN rates}
\label{localratio}

Mannucci et al. (2005) have reanalysed the SN survey data
described in Capellaro et al. (1999), to obtain core-collapse
and type-Ia SN rates for distinct galaxy types.  
They, and several subsequent studies (Scannapieco \& Bildsten 2005;
Mannucci et al. 2006; Sullivan et al. 2006),
have found that the SN-Ia
rate in star-forming galaxies 
is dominated by a ``prompt'' progenitor component. This component
has  a typical delay time between stellar formation 
and SN-Ia explosion of $\sim 10^8$~yr or
less. This leads to a linear dependence of
SN-Ia rate on the star formation rate of the host galaxy population.
In star-forming galaxies, Mannucci et al. (2005) find a ratio
of $N_{\rm Ia}/N_{\rm CC}=0.35\pm 0.08$ between the type-Ia and core-collapse
rates. 
This ratio permits a first estimate of the fraction of the 
potential type-Ia progenitor stellar
population that actually explodes through the prompt type-Ia channel:
\begin{equation}
\label{frac0}
\eta_{\rm Ia/CC}=\frac{N_{\rm Ia}}{N_{\rm CC}}
\frac{\int_8^{50} (dN/dm) dm} 
{\int_3^8 (dN/dm) dm} .
\end{equation} 
For the standard parameters, the ratio of the two integrals equals
0.332, giving  $\eta=0.12\pm 0.03$.
In the integral in the numerator, the upper mass limit is 50,
since more massive stars may collapse directly to black holes
without a SN explosion. Lowering the upper limit
to 25 reduces $\eta$ by 14\%, and raising it to 100 
increases it by 9\%. Including this uncertainty gives,
for the standard parameters, an estimate of $\eta_{\rm Ia/CC}=0.08-0.15$. 

\subsection{The SN-Ia rate per unit star-formation rate}
\label{Bmeas}

A ``prompt'' SN rate per unit star formation rate, $B$
(adopting the notation introduced by Scannapieco \& Bildsten 2005), has
dimensions of a SN number divided by a stellar mass. The value of $B$ equals
the total number of prompt SNe-Ia that have exploded among an
actively star-forming 
population since star formation
began, divided by the total stellar mass that has formed. 
Assuming that the population is being observed at least $\ga 10^8$~yr
after the onset of star formation, the fraction
of  stars with initial 
masses in the range $m_{\rm min}-m_{\rm max}$ 
that eventually explode as SNe-Ia is therefore 
\begin{equation}
\label{frac1}
\eta_{B}=B\frac{\int_{0.1}^{100} m (dN/dm) dm} 
{\int_3^8(dN/dm) dm}.
\end{equation} 
For the standard parameters, the ratio of the two integrals equals 47.6.

It is important 
to remember that SN rate measurements are not normalized by a directly
measured star formation rate. Rather, the luminosity and the 
color, or set of colors, of the host galaxy population are used to
derive a star formation rate, usually by comparison to spectral 
synthesis models. The luminosity and colors of a stellar
population at a given time is dominated by the most massive stars 
that are evolving off the main sequence at that time, while the mass
of the population is dominated by the least massive, but most
numerous, stars, The derivation of a star formation rate therefore 
requires the assumption
of an IMF for the largely unobserved low-mass population. 
For self-consistency, the right side
of Eq.~\ref{frac1} should then be multiplied by  a correction factor
\begin{equation} 
\label{imfcorr}
g_{\rm imf, org}=\frac{\int_{m_{\rm low,org}}^{m_{\rm to,org}}m (dN/dm)_{\rm org} dm}{\int_{m_{\rm low}}^{m_{\rm to}}m (dN/dm) dm},        
\end{equation}
where $m_{\rm to}$ is the turnoff mass, and the ``org'' subscript 
denotes the original IMF assumed in the published rate 
measurement. Previous studies that have derived SN rates have
generally assumed the same, or similar, IMF parameters as the standard
ones I have chosen, making $g_{\rm imf, org}\approx 1$. 
I will note, below, when this is not the case.   

Dahlen et al. (2004) measured the SN-Ia rate vs. redshift $z$, out to
$z\sim 1.6$. They took a parametric fit to 
the star-formation rate vs. cosmic time, $SFR(t)$,
found by Giavalisco et al. (2004), convolved it
with various parametrised 
trial SN-Ia delay time distributions, and fitted the observed  
SN-Ia rate vs. $z$. The normalization in this fit,
$(1.0-1.3)\times 10^{-3} M_{\odot}^{-1}$
 for the various delay functions, gave the
number of SNe-Ia per unit formed stellar mass. As in
Eq.~\ref{frac1} above, Dahlen et al. (2004) multiplied this by  the ratio
of the same two integrals (they used $m_{\rm hi}=125$, rather than
$m_{\rm hi}=100$, which makes a negligible difference), to obtain $\eta_B=5\%$
to $7\%$.  

Barris \& Tonry (2006) likewise compared the $SFR(t)$ of Giavalisco et
al. (2004) to the SN-Ia rates they measured themselves,
but used an interpolated form
of the $SFR(t)$, rather than a fitted functional form, as done by 
Dahlen et al. (2004). Furthermore, they limited their comparison 
to $z<1$, and rather than convolving the $SFR(t)$ with a delay function,
they shifted it by a set delay (equivalent to convolution with a
shifted $\delta$-function, which is reasonable for the short delays
they considered). They found a best-fit delay of $0-1$~Gyr, a
number of  SNe-Ia per stellar mass of $(1.7-2.2)\times 10^{-3} M_{\odot}^{-1}$,
and $\eta_B=8\%$ to $10\%$.  

Scanappieco and Bildsten (2005) used the Dahlen et al. (2004)
measurements at $z\le 1$ of type-Ia and core-collapse SN rates, and
the Giavalisco et al. (2004) $SFR(t)$, to derive  
$B=(2.6\pm 1.1)\times 10^{-3} {\rm SNe} {M_{\odot}}^{-1}$.
For the standard IMF parameters, this
gives $\eta_{B}=12\pm 6\%$. Alternatively, Scanappieco and Bildsten
(2005) multiplied the SN-Ia rate measured by Mannucci et al. (2005)
in starburst galaxies
 by the age of the stellar population of
these galaxies (as estimated based on their colors), to obtain  
$B=(1.2\pm 0.6)\times 10^{-3} {M_{\odot}}^{-1}$,
which I translate to $\eta_B=3\%$ to $9\%$.  

Sullivan et al. (2006) estimated star-formation rates, 
based on galaxy broad-band
spectral energy distributions, for the SN host galaxies 
in the Supernova Legacy Survey, at 
$0.2<z<0.75$. They found a prompt component of the rate
of  $B=(3.9\pm0.7)\times 10^{-4} 
{M_{\odot}}^{-1}$. To obtain their mass estimates, 
Sullivan et al. (2006) assumed a Kroupa (2001) IMF, which has a slope
of $\alpha=-2.3$ above a mass of $m=0.5$ and $\alpha=-1.3$ below it. If the 
Kroupa (2001) and Salpeter (1955) 
IMFs are both normalized at a mass of 8 (which corresponds
roughly to the B-type stars that dominate the light of young stellar
populations), the ratios of the respective total masses are 0.69. Converting 
from the Kroupa (2001) IMF to the
standard IMF parameters therefore increases the total stellar mass
formed, and hence reduces $B$ by 0.69. 
The measurement of $B$ by Sullivan et al. (2006), adapted to the
standard IMF parameters, then implies $\eta_B=1-1.5\%$

Mannucci et al. (2006) have tested a large variety of delay time
distributions,
some parametrized and some physically motivated, to fit simultaneously: 
(a) the SN-Ia rate vs. redshift (by convolving, again, the 
parametric fit to the $SFR(t)$,
found by Giavalisco et al. 2004, with the delay function); (b) the SN-Ia
rate per unit mass as a function of galaxy colors measured by Mannucci
et al. (2005); and (c) the dependence of SN-Ia  
rate on galaxy radio loudness found by Della Valle \& Panagia (2003)
and Della Valle et al. (2005). As before, the normalization of the 
fit to the SN-Ia rate vs. redshift gives $\eta$. For both of their
best-fitting models -- a physical SD model by
Belczynnski et al. (2005), and a parametrized two-component delay function,
consisting of equal contributions from a prompt component and from a delayed
component having an exponential decay with characteristic time 3~ Gyr
-- they found values of $\eta\approx 4\%$. 

Thus, for the standard parameters, these six estimates imply $\eta_B$ 
in the range $1-10\%$. This can be compared to $\eta_{\rm
  Ia/CC}=8-15\%$, found above, as both relate to the same prompt
SN-Ia component. 

\subsection{The SN-Ia rate per unit stellar mass}

A third estimate of $\eta$ can be obtained from the ``tardy''
component of the SN-Ia rate identified by Mannucci et al. (2005),
and which dominates in old stellar populations, whose ages are 
 $\sim 10$~Gyr.
This is effectively a ``DC'' component in the SN rate, whose level
is proportional to the total stellar mass, and is denoted by $A$
(again following Scannapieco \& Bildsten 2005).
 Assuming that the maximum delay
time between star formation and SN-Ia explosion for this component is 
$\sim 10$~Gyr, then $A\times 10$~Gyr is the number of SNe-Ia
per unit stellar mass that explode through this channel, and 
the SN-Ia explosion fraction in this case is
\begin{equation}
\label{frac2}
\eta_{A}=
A\times 10^{10}{\rm yr}\frac{\int_{0.1}^{100}m (dN/dm) dm} 
{\int_{3}^{8}(dN/dm) dm}.
\end{equation} 

Formally, the integral in the numerator should run up only to 
the main-sequence turnoff mass in old stellar populations, $m\approx
1$. Above $m\approx 1$, the numerator should include only the
contributions from stellar remnants -- WDs, neutron stars, and black
holes. More to the point, like Eq.~\ref{frac1}, 
Eq.~\ref{frac2} needs to be corrected
by the factor $g_{\rm imf, org}$ (Eq.~\ref{imfcorr})
appropriate for the IMF assumed in the original SN-Ia
rate-per-unit-mass calculation,
 when deriving the mass of the stellar
population from its observed luminosity and colors. Often, 
the total formed mass, rather than the mass in stars and
in remnants, has been used. 

Note that Eq.~\ref{frac2} assumes that, $\sim 10$~Gyrs
after the formation of the stellar population, we are
seeing the last SNe-Ia that form through the tardy channel. If the 
SNe, in reality, continue to explode at a constant rate for, say, 20~Gyr (i.e.,
the supply of SN-Ia progenitors has been only half used up after
10~Gyr), then $\eta$ would be doubled. Alternatively, if the tardy
component is not constant, but, say, decays exponentially
with a timescale of 10~Gyrs, then Eq.~\ref{frac2} is exact.

Mannucci et al. (2005) derived a local value of the mass-normalized
SN-Ia rate in early-type galaxies of $A=(3.8\pm 1.3)\times 10^{-4} {\rm yr}^{-1} ({10^{10} M_{\odot}})^{-1}$ (after converting their
measurement from $H_0$ of $75$ to $70$~km~s$^{-1}$Mpc$^{-1}$). 
Their mass determination 
was based on the calculations of Bell \& de Jong (2001), who defined
relations between the mass-to-light ratios and the 
optical-to-near-IR colors of synthetic galaxies. Bell \& de Jong
(2001) assumed a Salpeter (1955) IMF, but scaled down by a factor
$0.7$, which they denoted a ``diet'' Salpeter IMF.   
Including this factor in Eq.~\ref{frac2}, we obtain $\eta_A=0.8-1.7\%$

Sullivan et al. (2006) found, based on the early-type galaxies
in the Supernova Legacy Survey, 
a tardy component of the SN-Ia rate of $A=(5.3\pm 1.1)\times 10^{-4} {\rm
 yr}^{-1}
({10^{10} M_{\odot}})^{-1}$. As noted above, in obtaining their mass
estimates from stellar synthesis modeling, they assumed a Kroupa
(2001) IMF, and only the mass in surviving stars.
If the Salpeter (1955) 
and Kroupa (2001) IMFs are both normalized at a mass of 1 (which corresponds
roughly to stars that dominate the light of old stellar
populations), the ratio of the total stellar masses below $m=1$ 
equals 2.0. Converting 
from the Kroupa (2001) IMF to the
standard IMF parameters therefore increases the total stellar mass
formed (and hence reduces $A$) by a factor of 2, giving $\eta_A=1-1.5\%$. 

Sharon et al. (2007) have recently measured a SN-Ia
rate in massive clusters at $z\approx 0.1$ (based on the SN survey by
Gal-Yam et al. 2007), 
of $A=9.8^{+6.8}_{-4.8}\times 10^{-4} {\rm
yr}^{-1} (10^{10} M_{\odot})^{-1}$. Since there is little star
formation in clusters at these redshifts, this is effectively another 
measurement of the tardy $A$ component of the SN-Ia rate.
Following Mannucci et al. (2005), the mass determination 
was based on the mass-to-light ratios found by Bell et al. (2003), 
who assumed a Salpeter (1955) IMF, scaled down by a factor $0.7$.   
Including this factor in Eq.~\ref{frac2}, we obtain $\eta_A=2-6\%$

Mannucci et al. (2007), again using the Cappellaro et
al. (1999) sample, find a rate in local early-type cluster galaxies
of $A=6.6^{+2.7}_{-2.0}\times 10^{-4} {\rm
yr}^{-1} (10^{10} M_{\odot})^{-1}$, corresponding to $\eta_A=2-3.5\%$ 

We thus obtain values of 
$\eta_{A}$ between 1\% and 6\% for the tardy component.
The total fraction of the initial $m=3-8$ stars in a stellar population that
explode as SNe-Ia through either one of the prompt or tardy channels is  
$\eta=\eta_{ A}+\eta_{B}$ or $\eta=\eta_{ A}+\eta_{\rm Ia/CC}$. 
Taking the extreme values found above, this gives a range for $\eta$ of 
2\% to 21\% for the standard parameters.

\subsection{The iron to stellar mass ratio in galaxy clusters}

The X-ray spectra of massive galaxy clusters reveal a large mass of 
iron, most of it in the hot intra-cluster medium (ICM, e.g., Balestra
et al. 2007, and references therein).
A tight correlation is observed between the iron mass and the 
total luminosities of the early-type galaxies in clusters, while no
correlation is observed with the sub-dominant late-type galaxy luminosity.
This strongly suggests that the source of the iron is SNe from 
the stellar 
population whose low-mass component currently constitutes the early-type
galaxies (rather than, say, the late-type galaxies, or infall of
external material into the cluster). Assuming the standard IMF, it can
be shown (e.g., Maoz \& Gal-Yam 2004) 
that core-collapse SNe could have produced only about 20\% of
the total mass of iron in the ICM and in the galaxies themselves. If the
remaining 80\% of the iron was formed by SNe-Ia, then the ratio of this 
iron mass to the current stellar mass gives another, independent, estimate
of $\eta$:
\begin{equation}
\label{frac3}
\eta_{\rm Fe}=\frac{0.8 M_{\rm Fe}}{Y_{\rm Fe}M_*}
\frac{\int_{0.1}^{100}m (dN/dm) dm} 
{\int_{3}^{8}(dN/dm) dm},
\end{equation}
where $M_{\rm Fe}/M_*$ is the ratio of total iron mass to current
total stellar mass, and $Y_{\rm Fe}$ is the mean iron yield of a
single SN-Ia.
 The first term in the product on the right side is
just the cumulative number of SNe-Ia that have exploded in
the cluster, divided by the stellar mass. 

 From analysis of X-ray, optical, and IR measurements, Lin, Mohr, \&
 Stanford (2003) have estimated the iron-to-stellar mass ratio of clusters.
Their figure 9, with their assumed Solar iron abundance of
$Z_{\odot}=1.814\times 10^{-3}$ (Anders \& Grevese 1989), suggests
$M_{\rm Fe}/M_*=(3\pm 0.5)Z_{\odot}=(5.4\pm 0.9)\times 10^{-3}$. 
Lin et al. (2003) found the stellar
masses in their sample clusters based on the 2MASS survey $K$-band 
luminosities of the cluster galaxies. The spiral galaxy stellar 
mass-to-light ratio, $M/L$,
was obtained from the relations with
luminosity and color  of Bell \& de Jong (2001), who assumed  the
 ``diet'' Salpeter IMF (i.e., scaled down by a factor 0.7, see above). For the 
ellipticals, Lin et al. (2003) used the dynamical estimates by Gerhard
 et al. (2001) of $M/L$ as a function of luminosity in a sample of 21
ellipticals. Lin et al. (2003) found, for the clusters as a whole,
a range of $K$-band $M/L$ between 0.7 and 0.8, in solar units. They
 noted the similarity of this value to the mean value of 0.73 obtained by Cole
et al. (2001) for the 2dF galaxy redshift survey. In deriving this
value, Cole et al. had assumed a Kennicutt (1983) IMF, which has a
 slope of $\alpha=-1.4$ for $m<1$ and $\alpha=-2.5$ at $m>1$. Converting from a
 Kennicutt (1983) to a Salpeter (1955) IMF increases the $K$-band
 $M/L$ of an old stellar population by a factor of 2.0 (Bell et al.
 2003). Thus, for our 
standard IMF parameters, the Lin et al. (2003) iron-to-stellar mass ratio
is reduced by a factor of 2, to   
$M_{\rm Fe}/M_*=(2.7\pm 0.5)\times 10^{-3}$. 

The mean iron yield of a SN-Ia is 
observationally constrained by SN-Ia light curves and spectra,
as well as from SN~Ia model calculations, 
to be $Y_{\rm Fe}\approx 0.7 \pm 0.1 M_{\odot}$
(e.g., Thielemann, Nomoto, \& Yokoi 1986; Nugent et al. 1997; 
Contardo, Leibundgut, \& Vacca 2000; Mazzali et al. 2007; although 
the iron mass yield among individual SNe-Ia 
likely has a dispersion of a factor of a few, for the present estimate we are
actually interested only in the mean, and the error on the mean). 
Accounting for the uncertainties then
gives a range $\eta_{Fe}=11-20\%$. 

The core-collapse contribution to the 
iron production may have been greater than the 20\% stated above, 
if the first stellar generations formed with a top-heavy IMF that 
favored massive stars that exploded as  core-collapse SNe, or if 
many of the ancient core-collapse SNe were more efficient iron
producers than present-day core-collapse SNe (see Maoz \& Gal-Yam 2004;
Loewenstein 2006). Either of these
non-standard scenarios would lower $\eta_{Fe}$. On the other hand,
Kapferer et al. (2007) find, by performing simulated X-ray
observations of simulated clusters, that X-ray measurements
systematically underestimate the mass of metals in the ICM. Correcting
for this effect would raise $\eta_{Fe}$.

The low SN-Ia rates observed in clusters
at $z<1$ (Gal-Yam et al. 2002; Sharon et al. 2007; Mannucci et
al. 2007) suggest that only the tardy SN-Ia channel is presently
active, while much of the iron was produced at early times
by a prompt channel.
The iron mass in clusters is likely a tracer of the cumulative
products of both the prompt and tardy SN-Ia
channels. If so, the range of $\eta_{Fe}$, 11-20\%, 
  should be compared to $\eta_A+\eta_B$,
or $\eta_A+\eta_{\rm Ia/CC}$, above, which have a range of $2-21\%$. 

\subsection{The abundance-based ratio of core-collapse 
to type-Ia SNe in clusters}

The large effective areas and high X-ray spectral resolutions of {\it
  Chandra} and {\it Newton-XMM} have permitted
accurate measurements of the ICM abundances of additional elements,
beyond iron. De Plaa et al. (2007) have recently derived abundances
for a sample of clusters, and fitted to the abundance patterns the 
theoretical yields of core-collapse and type-Ia SNe. They then derived the
relative numbers of the two types of SNe, whose cumulative outputs
have contributed to the observed ICM abundances. Their best fit uses,
for the SN-Ia contribution, 
a delayed detonation model by Badenes et al. (2003) 
that Badenes et al. (2006) had found
  provides the best fit to the X-ray emission from the Tycho supernova 
remnant. The best fit to the ICM abundances then  
indicates that the number ratio of the two types of SNe that have 
exploded in clusters
  is in the range $N_{Ia}/N_{cc}=0.5-1.2$.\footnote{This result is not in
  conflict with the previous argument, that 80\% of the iron in
  clusters may come from SNe-Ia. Each SN-Ia produces an
  order-of-magnitude 
  more iron than a core-collapse SNe, so both results suggest that 
  SNe-Ia dominate the iron
  production. However, one must keep in mind that the elemental yield
of core-collapse SNe based upon explosion models
   is extremely uncertain.} This leads to a fifth estimate of 
 $\eta$:
\begin{equation}
\label{frac4}
\eta_{\rm abund}=\frac{N_{Ia}}{N_{cc}}
\frac{\int_8^{50} (dN/dm) dm} 
{\int_3^8(dN/dm) dm} , 
\end{equation} 
which is the same kind of calculation as in Eq.~\ref{frac0}, above.
De Plaa et al. (2007) performed this calculation, based on their
derived $N_{Ia}/N_{cc}$ ratio, and the same assumed IMF parameters,
but taking $m_{\rm min}=1.5$ or 0.9 rather than 3, $m_{\rm max}=8$ or
10, and either a Salpeter or a Kroupa (2001) IMF. Using $m_{\rm min}=3$,
and again allowing also for the uncertainty in the upper mass limit
for core-collapse SNe, I find $\eta_{\rm abund}=14-40\%$. 
Like $\eta_{\rm Fe}$ above, $\eta_{\rm abund}$ measures the
exploding fraction through both SN-Ia channels. 

\subsection{Dependence of $\eta$ on assumed parameters}
\label{imfvary}

For the standard parameters with a Salpeter IMF (i.e., a single power
law of index $\alpha=-2.35$), the integrals above are trivially solvable,
and give the stated $\eta$ fractions. I now investigate briefly the 
effect or varying those parameters, within plausible values. 

Most of the forms that have been proposed for the IMF differ mainly
at $m<1$, while at $m>1$ they are usually single power laws of similar
slopes, between $-2.2$ and $-2.7$ (e.g., Baldry \& Glazebrook 2003,
$\alpha=-2.2$; Kroupa 2001, $\alpha=-2.3$;
Chabrier 2003, $\alpha=-2.3$; Salpeter 1955, $\alpha=-2.35$; Gould et al. 1997 $\alpha=-2.35$;
Kennicutt 1983, $\alpha=-2.5$; Kroupa et al. 1993, $\alpha=-2.7$; an exception is the Scalo 1998 IMF, which
has $\alpha=-2.7$ for $1<m<10$ and $\alpha=-2.3$ at $m>10$). 
The actual observables in SN rate
measurements are the ratios of SN numbers of different types
(Ia to core-collapse), or of a SN number and a stellar luminosity, 
the latter dominated
either by $m\approx 1$ stars in old populations, or by $m\approx 8$
stars in young populations. Normalizations of SN rates
 relative to total stellar
mass or total star formation rate are always obtained by assuming
some IMF over the entire stellar mass range. However, estimates of
$\eta$ relate only to the stellar populations in the intermediate mass
(SN-Ia progenitors) and high mass (core-collapse progenitors) ranges.
In my estimates of $\eta$, above, I have 
taken account of the IMFs that were assumed in the original
rate calculations and converted them consistently to estimates based on
the standard IMF parameters. Any actual dependence of $\eta$ 
on the IMF will therefore be only through the form of the IMF at $m>1$.   

 IMFs that are flatter in the 
$m_{\rm min}-m_{\rm max}$ range will yield either fewer or 
more progenitors, depending on whether the IMF is normalized
at the high end or the low end, respectively. 
For $\eta_A$ and $\eta_{\rm Fe}$,
the IMF normalization needs
to be made at $m\approx 1$, the stellar mass that dominates the light of
the old populations whose mass normalizes those SN-Ia number
estimates. Therefore, a flatter IMF slope
will increase the number of SN-Ia progenitors and lower $\eta$, 
and vice versa. Conversely, $\eta_B$ is based on SN-Ia rates 
normalized by the light of young stellar populations, dominated
by the O- and B-type stars at the high end of the SN-Ia progenitor range.
A flatter IMF slope will raise $\eta_B$, and vice versa.
The estimates of $\eta_{\rm Ia/CC}$ and $\eta_{\rm abund}$ are based
on number ratios of high mass and intermediate-mass stars. A flatter IMF slope
will always increase the number of high-mass stars relative to
low-mass stars, and therefore $\eta$ will increase in these cases.  

\begin{figure}
\includegraphics[width=0.5\textwidth]{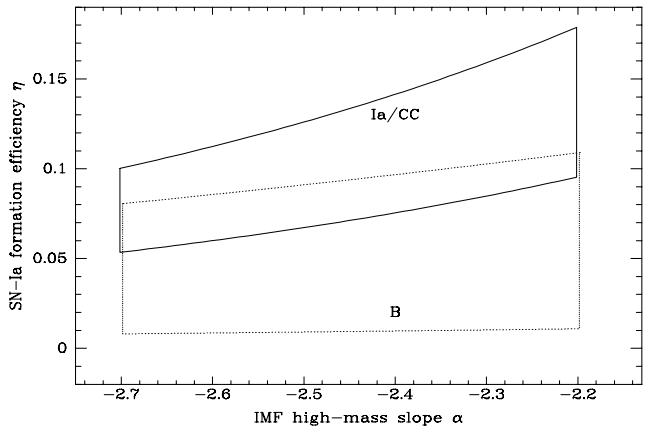}
\caption{Allowed regions in the parameter space of $\eta$, the
  fraction of all stars with ZAMS mass $m=3-8$ that explode as SNe-Ia,
and $\alpha$, the power-law slope of the IMF at $m>1$. The two regions
shown are based on two methods for estimating the fraction of stars that
explode only through a ``prompt'' SN-Ia channel, that occurs within
$\sim 10^8$~yr of star formation. $\eta_B$ (dotted curve) is based on
measurement of the SN-Ia rate per unit star-formation rate in
  star-forming galaxies. $\eta_{\rm Ia/CC}$ is based on the ratio of
Type-Ia and core-collapse SN rates in local star-forming galaxies.
The vertical extent for each method is based on the union of the
 $1\sigma$ uncertainty ranges of the different measurements, described
in the text. The left and right borders encompass the range
of high-mass slopes of most standard IMFs. 
}
\label{fig1}
\end{figure}

\begin{figure}
\includegraphics[width=0.5\textwidth]{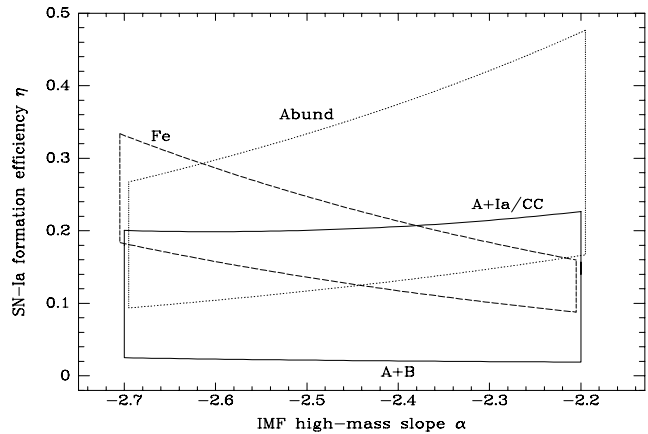}
\caption{Same as Fig.~1, but for estimates of $\eta$ that reflect the
{\it total} fraction of intermediate-mass stars that explode as
SNe-Ia, through either a prompt or a tardy channel. The dashed curve,
labeled Fe, is based on the iron-to-stellar mass ratio
in galaxy clusters. The dotted curve, for $\eta_{\rm abund}$, is based 
on the deduced ratio of Ia to core-collapse SN numbers that have
contributed to observed ICM abundances. The lower
border of the solid curve is the sum of $\eta_B$ (shown in Fig.~1),
and $\eta_A$, the fraction exploding through the tardy component,
based on SN-Ia rates in old populations. The upper border of the solid
curve is the sum of  $\eta_{\rm Ia/CC}$, shown in Fig.~1, and $\eta_A$.    
A value of $\eta\approx 15$\% is consistent with all of the methods
and most of the IMF slopes.
}
\label{fig2}
\end{figure}

Figure 1 shows the ranges of the two estimates of $\eta$ for the prompt 
channel, namely  $\eta_{\rm Ia/CC}$ and $\eta_B$, and their dependences
on the IMF slope, $\alpha$, above $m=1$. For each method, at each
value of the IMF slope, the range of permitted $\eta$ is the union of
the $1\sigma$ uncertainty ranges of the different measurements based on the
method. There is a broad range of
consistency between these two estimates in the range $\eta=5$ to
10\%, for the various plausible high-mass IMF slopes. Note that $\eta_{\rm
  Ia/CC}$, which is proportional to the observed ratio of type-Ia 
and core-collapse rates, will be systematically biased high if an 
unaccounted fraction of core-collapse SNe in star-forming galaxies
are undetected because
they explode in highly obscured regions. Indeed, a number of studies
(e.g., Maiolino et al. 2002; Mannucci et al. 2003; Cresci et al. 2007)
have concluded that a large fraction of core-collapse SNe in starburst
galaxies are missed by optical surveys. Mannucci et al. (2005) show,
based on a simple model, that of order one-half of the local 
core-collapse SNe could possibly be missing. Correcting for this
would lower $\eta_{\rm Ia/CC}$ by one-half and result in a larger overlap
in Fig.~1 between the allowed areas corresponding to $\eta_B$ and $\eta_{\rm
  Ia/CC}$. A further reduction in $\eta_{\rm Ia/CC}$ will result if
$m=8-10$ stars form ONeMg WDs, rather than
core-collapse SNe, which is still the subject of theoretical uncertainty
(see, e.g., review in Herwig 2005), or 
if low-mass or low-metallicity core-collapse progenitors produce
faint SNe that are missed by surveys. For example, limiting the
detectable core-collapse SNe to those with initial $m>12$ would lower 
$\eta_{\rm Ia/CC}$ by another factor of $\sim 0.5$.  
  
Figure 2 is the same as Fig.~1, but for the estimates of $\eta$ that trace
the cumulative effect of both the prompt and the delayed channels,
i.e., $\eta_A+\eta_B$, $\eta_A+\eta_{\rm Ia/CC}$,
$\eta_{\rm Fe}$, and $\eta_{\rm abund}$. For clarity, I have plotted
the union of the regions in parameter space allowed for 
$\eta_A+\eta_B$ (lower limit of the region) and $\eta_A+\eta_{\rm
  Ia/CC}$ (upper limit of the region).
Interestingly, a value of 
$\approx 15\%$ is consistent with all of these 
different methods of estimating the total $\eta$, as long as $\alpha$ 
assumes one of its conventional values. 

In terms of varying the other standard parameters,
enlarging the range $m_{\rm min}-m_{\rm max}$ will naturally
increase the number of potential SN-Ia progenitors, and therefore
will lower $\eta$. Estimates of $\eta$ that involve also the
core-collapse SN rate will be further lowered by the smaller range
of core-collapse progenitors resulting 
when $m_{\rm max}$ is increased. 
For a Salpeter IMF, lowering $m_{\rm min}$ from 3 to 2.5, 2, or 1.5
lowers $\eta$ by factors of 0.72, 0.5, or 0.32, respectively.
For estimates of $\eta$ not involving the core-collapse rate, 
raising $m_{\rm max}$ to 9, 10, or 11
lowers $\eta$ by another factor of 0.94, 0.91, or 0.89, respectively.
When the core-collapse rate is involved ($\eta_{\rm Ia/CC}$ and
$\eta_{\rm abund}$), the respective reductions are 0.80, 0.65, and 0.55.
On the other hand, there is ongoing uncertainty regarding the highest
initial stellar mass that leads to a CO WD. Reviewing the 
stellar-evolution models and the observational evidence (the initial-final
stellar mass relation), Weidemann (2000) argues for a 
highest initial mass of $m=6-7$, although some models (Girardi et
al. 2000) find even lower limits, of $m\sim 5$. A shrinking
of the $m=3-8$ progenitor range to $m=3-5$ or $m=3-6$ 
will naturally raise $\eta$. All 
these changes can thus amount to reductions or increases
by up to a factor of a few in all of the estimates of $\eta$, derived above.

\section{Binarity, mass ratio, separation, and duplicity}

The fraction $\eta$, discussed above, is the fraction of 
all stars with ZAMS masses in the prescribed range that explode as
SNe-Ia. To estimate the fraction of all {\it close binary systems}
with members in this mass range that lead to a SN-Ia explosion, 
one most consider the binarity fraction, the binary mass-ratio distribution,
the separation distribution, and the duplicity of the progenitors.

%If a fraction $f$ of stars are in binary
%systems and $1-f$ are single, the number of binary systems
%relative to the total number of stars is  $f/2$.
The binarity fraction among stars is a function of mass, with 
most low-mass stars being in single systems, while most massive stars are 
members of binaries 
(see Lada 2006, and references therein).
Amongst the intermediate-mass population of
interest here (B-type and A-type stars), 
the binarity fraction (defined as  the fraction
of systems that are binary, rather than single-star, systems)
is likely in the range $b\approx 1/2$ to $1$ (e.g., Shatsky \& Tokovinin 2002;
Kouwenhoven et al. 2005; Baines et al. 2006).
%De Marco et al. (2004) have also found, based on radial velocity
%measurements of planetary nebula nuclei, that a large fraction
%of planetaries may have binary nuclei with periods of up to a few months.  
If a fraction $b$ of systems are binaries, and $1-b$ are single,
the fraction of all stars that is in binaries is $f_a=2b/(b+1)$.
Thus, $f_a\approx 2/3$ to $1$.

Several recent studies have measured 
the binary mass ratio distribution, $f(q)$, where $0<q<1$ is the
ratio of secondary to primary mass, for intermediate-mass stars. 
For A-type stars, Soederhjelm (2001) finds that
$f(q)\sim {\rm const.}$, for separations of $60-120$~AU.
For B-type stars, Shatsky\& Tokovinin (2002) find $f(q)\sim q^{-0.5}$
to $q^{-0.3}$
at separations of $45-900$~AU. 
Kouwenhoven et al. (2005) measure $q^{-0.33}$ at $30-1600$~AU for 
A-type and late B-type stars in the Sco OB2 association.
A distribution $f(q)$ that is flat
or slowly rising toward small $q$ appears to be generic for a wide 
range of masses and separations. For example, Mazeh et al. (2003)
find such an $f(q)$ for lower-mass (late G and
early K) close binaries with periods $<100$~days (i.e., separations
$\la 1$~AU).
The fraction of primaries with $m=3-8$
that have a secondary also in this mass range (as required in the DD
scenario) is
\begin{equation}
\label{massratio}
f_b=
\frac{\int_3^8 (dN/dm) dm \int_{q=3/m}^1 f(q) dq} 
{\int_3^8(dN/dm) dm}.
\end{equation} 
Assuming $f(q)\propto q^{-0.5}$ (and normalizing $f(q)$ so that its
integral from 0 to 1 is unity), we find $f_b=0.17$.
Taking $f(q)={\rm constant}$ gives $f_b=0.29$.
Thus $f_b\approx 1/6$ to 1/3. The fraction $f_b$ depends only weakly on the IMF slope
in the $m=3-8$ range.

The initial distribution in physical separation in binaries, $a$, 
is still debated. Oepik (1924) already suggested
that this distribution has the form $f(a)\sim a^{-1}$, but 
other forms have been suggested (e.g., Duquennoy \& Mayor 1991). 
Poveda et al. (2006) have recently analyzed several samples 
of binaries, and concluded that the Oepik form provides an excellent
fit over a very wide range, from $a=60$ to 45,000~AU (the upper limit
determined for binaries in the young Orion Nebula cluster,
which have not had time to be disrupted by encounters
with massive perturbers), and suggesting
that the same physical process operates at all scales.  
Both of the currently
popular scenarios for SNe-Ia, SD and DD,
require passage of the progenitor binary through at least one common-envelope
phase. For initial $m=3-8$, the maximum radii that single stars achieve
during their post-main-sequence evolutions are 2-5~AU (see, e.g.,
fig.~1 of Yungelson 2005). This dictates an initial separation 
$\la 2-5$~AU, if the binary is to undergo common-envelope evolution. 
Taking the full 
range of possible initial separations from $2\times 10^{-2}$~AU 
(contact between two $m=3$ ZAMS stars) to $10^{4.7}$~AU, 
and assuming the $1/a$ distribution,
a fraction of $f_c\approx 0.3-0.4$ of binaries have an initial
separation within 2-5~AU. If we assume, ``generously'', that even
binaries with initial $a\sim 50$~AU can migrate to smaller separations
(e.g., via interactions with protoplanetary disks), then $f_c\approx
1/2$. Thus, $f_c$ is likely in the range 1/4 to 1/2, but probably 
at its low end. 

Finally, 
every DD binary can lead, at most, to
one SN-Ia, and therefore the duplicity factor is $f_d=1/2$.

If $\eta$ is the fraction of all intermediate-mass stars
that explode as SNe-Ia, the fraction of all intermediate-mass 
close binaries that explode is $f_{Ia}=\eta/(f_a f_b f_c f_d)$.
The range formed by the products of the extreme values of $f_a$ through $f_d$,
above, gives $f_{Ia}\sim (12-70) \eta$.
{\it For all of the observationally derived 
values of $\eta$ derived above, a fraction
of order 100\% of all intermediate-mass
close binaries explode as SNe-Ia.} For some estimates of $\eta$ and 
of the binary parameters (for example, for
$\eta=15\%$, a value that is consistent with all methods, and a
product of the binary parameters of 70), 
there are up to an order of magnitude more
SNe-Ia than potential progenitor systems .
  
\section{Discussion}
\label{discussion}
The empirical derivation of a high SN-Ia formation efficiency is not 
new. For example, based on observations by Tammann (1978) that the type-Ia and
core-collapse SN rates in spiral galaxies are equal, 
Greggio \& Renzini (1983) already reached the same
conclusion. 
The authors of several of the
observational studies whose results I have used above have themselves
made the same point. The novelty in the present work is
that: (a) I have based my estimate of $\eta$ on a number of 
relatively accurate measurements of extragalactic SN rates, and of related
observables, that are recently available; (b) I have compared among 
the values of 
$\eta$ and its uncertainty resulting from different measurements of the same
type (e.g., SN Ia-rate per unit star-formation rate), and to those
derived by independent methods, all after correcting  
for the different IMF assumptions that went into the various
estimates; (c) I have tested for the dependence of the results
 on IMF assumptions, and argued that they depend only on the high-mass
($m>1$) slope, which is fairly well constrained.
(d) For all of these estimates, I have incorporated recent results on binarity
fraction, binary separation, and binary mass-ratio distribution, to 
estimate the fraction of close, intermediate-mass, binaries that explode.  

I have shown that, if our basic ideas about SN-Ia progenitors
are correct, then, at best, the processes leading to a SN-Ia explosion
are extremely efficient, in the sense that most or all of the
potential progenitor systems -- close, intermediate-mass, binaries --
do explode as SNe-Ia. At worst, some combinations of parameters imply
the existence of many more explosions than progenitor systems,
indicating an error in one or more of the assumptions. 
  
\subsection{Theoretical expectations}
What are the theoretical expectations for $f_{Ia}$? 
Calculations in which 
binaries with a range of initial parameters
are followed throughout their stellar and binary
 evolution, have been carried out by, e.g., 
Iben \& Tutukov (1984),
Hachisu et al. (1996, 1999a,b), Yungelson \&
 Livio (1998, 2000), Langer et al. (2000), 
Tutukov \& Yungelson (2002), Hurley et al. (2002), Han \& Podsiadlowski
 (2004), Fedorova et al. (2004), and Belczynski (2005). 
As summarized in Yungelson (2005), Tutukov \& Yungelson quantified
the fraction of all close intermediate-mass binaries that reach
the various binary evolution endpoints, including SN-Ia. They found
that the dominant channel for SNe-Ia is the DD merger channel,
with 0.7\%  of all binaries with both ZAMS masses in the $m=0.8-11.8$
range (the full range of masses that produce WDs within a Hubble time)
leading to a super-Chandra merger. Limiting the progenitor population
to $m=3-8$, the predicted fraction is $f_{Ia}=0.14$ (L. Yungelson,
private communication), considerably less than the $\sim 100\%$ 
I have derived from the observations. The SD channel was found to
be an order of magnitude less efficient in producing SNe-Ia. 

Hachisu et al. (1996, 1999a,b; see also Kobayashi et al. 1998, 2000) 
have argued against DD mergers as SN-Ia progenitors, and have
advocated, instead, two different, co-existing, SD paths to SN-Ia explosion:
a primary C/O WD with ZAMS mass $m=5-8$, accreting from a red-giant
secondary of initial mass $m=0.8-1.5$; or a primary of ZAMS mass 
$m=5.5-8.5$ that accretes from a ZAMS $m=1.8-3.4$ main sequence 
secondary.\footnote{Incidentally, Kobayashi et al. (2000) 
fit the element abundance
patterns in the Solar neighborhood using the yields from their SD
SN-Ia models and from core-collapse SNe. Their best fit value for the
total $\eta$, through both SD channels, is 7\%, similar to the
other estimates of $\eta$, above. This can be considered a sixth 
independent estimate of $\eta$, albeit a more model-dependent one.}  
Hachisu et al. claim that these SD channels are an order of magnitude
more efficient than estimated by Yungelson \& Livio (1998), i.e.,
comparable to the predictions for the DD channel favored by 
Tutukov \& Yungelson (2002), given above. 

Han \& Podsiadlowski 
(2004) have calculated more detailed numerical models for the 
WD + main sequence
SD channel. They find a similar, though somewhat smaller, range
of secondary initial masses, $m\approx 2-3.4$, that lead to an
explosion. In terms of rates, they find that this SD channel's
efficiency is intermediate to that in the high results of 
Hachisu et al. (1999a) and the low results of Yungelson \& Livio
(1998; see also Fedorova et al. 2004; 
see Han \& Podsiadlowski 2004, for 
a discussion of the sources of disagreement between the various studies.) 
The first SD channel, with a low-mass red giant secondary, was found
to contribute negligibly, in agreement with Yungelson \& Livio (1998). 

In contrast, Hurley et al. (2002; see also Tout 2005) found very 
low SN-Ia rates from their DD channels but rates comparable to the 
higher rates cited above for some of their SD channels. These authors
have also argued that $\alpha_{\rm CE}$, the parameter used to
represent in a simplified way the effects of the physically complex
common envelope stage of binary evolution, can
assume values larger than unity. The constant $\alpha_{\rm CE}$ is
the fraction of the orbital energy that goes into driving off the
envelope (see Nelemans \& Tout 2005, for an alternative parametrized
description of the common envelope stage). 
Most studies have assumed $\alpha_{\rm CE}=1$, but 
these authors have calculated also models with $\alpha_{\rm CE}=3$
which, they reason, are not unphysical, because additional sources
of energy (thermal, from nuclear burning, or magnetic, from dynamo
action) may be available, beyond the energy from rotation.
Tout (2005) emphasizes that, by using high values of $\alpha_{\rm
  CE}$, SN-Ia rates that are almost arbitrarily high  can be obtained.    

To compare the observational estimates 
also to the SD theoretical predictions, we need to take into account the
different progenitor mass ranges for this model -- instead of $m=3-8$
for both of the binary components, which I have used until now, we need to take
for the primary $m=5-8$, and for the secondary $m=0.8-3.4$ (``generously''
including also secondaries in the $m=1.5-2$ range). Due to the
smaller integration interval in the denominator, the   
$\eta$ estimates in Eqns.~\ref{frac0}-\ref{frac4} will grow by a
factor 3.1, for the standard IMF slope. In other words, the fractions of all
$m=5-8$ stars that explode as SNe-Ia will be  3 times higher than the
previous estimates. On the other hand, to describe the larger pool of 
secondaries, Eq.~\ref{massratio} needs to be replaced by
\begin{equation}
\label{massratiosd}
f_{b,\rm SD} =
\frac{\int_5^8 (dN/dm) dm \int_{q=0.8/m}^{3.4/m} f(q) dq} 
{\int_5^8(dN/dm) dm}.
\end{equation} 
Compared to the previous result, $f_b\approx 1/3-1/6$, this gives
$f_{b,\rm SD}\approx 1/2.5$, weakly dependent on the power-law slope of $f(q)$.
Finally, because the mass ranges of the primaries and the secondaries
are disjoint in this case, the duplicity factor, which we took as
$f_d=1/2$ for the DD case, is now 1. Thus, among all the potential SD
progenitor systems in the model of Hachisu et al. (1996, 1999a,b) -- close
binaries with a primary in the $m=5-8$ range -- the observed fraction of
systems that produce SNe-Ia will still be 0.4-0.8 of the estimate
obtained for the DD case, i.e., of order 100\%, or even more.
If we adopt the Han \& Podsiadlowski (2004) conclusion, that only $m=2-3.4$
secondaries can contribute non-negligibly to the SD channel, 
then $f_{b,\rm SD}\approx 1/4.3$, and we come even closer to the
DD result.

Comparing the observational and theoretical results, the typical 
observed value of 
$f_{Ia}\approx 100\%$ is an order magnitude larger than the
 DD prediction, and two orders larger than the
 the SD prediction of Tutukov \& Yungelson (2002).
The SD rate predictions of Hachisu et al. (1999a,b) and Tout (2005,
for $\alpha_{\rm CE}$=1) are comparable to the
the DD rate prediction of Tutukov \& Yungelson (2002), and the
SD predictions of Han \& Podsiadlowski (2004) are lower
than those of Hachisu et al. (1999a,b). Thus, all of these predictions
are lower by an order of magnitude, or more, than the observed
fraction. In fact, from fig. 6 of 
Han \& Podsiadlowski (2004), one can deduce\footnote{Like all of the above binary synthesis works, 
Han \& Podsiadlowski (2004) followed Iben \& Tutukov
(1984), in assuming a Galactic star-formation rate of 1 star
yr$^{-1}$ with $m>0.8$. Assuming our standard IMF, this translates to $5.8
M_{\odot}{\rm yr}^{-1}$, and gives the stated result for $B$.} 
a predicted value
of $B=(1-2)\times 10^{-4}~{\rm SNe} {M_{\odot}}^{-1}$, an order 
of magnitude lower than most of the measurements compiled in
\S\ref{Bmeas}.  
Even the lowest oberved value of $\eta$
found above, $\sim 2\%$ (which is inconsistent with all other
measurements), times the lowest values for 
$(f_a f_b f_c f_d)^{-1}\approx 12$, still leads to $f_{Ia}\approx
24\%$, twice
 as large as the highest theoretical predictions. 
 
Previous comparisons of theory and observation, in the binary synthesis
studies cited above, have focused on reproducing the Galactic SN-Ia rate,
after assuming a Galactic star formation rate. Both of these rates
are difficult to estimate, and are therefore highly uncertain. As a 
result, a factor-few agreement was considered satisfactory by those 
studies. The modern measurements of extragalactic 
SN rates, and related observables,
that I have used here make the 
order-of-magnitude discrepancy between theory and observations explicit. 

\subsection{Possible solutions}
\subsubsection{Observational errors}
To explain the discrepancy, we can re-examine some of the assumptions
in either the observational derivation or in the theoretical prediction.
First, we can consider the possibility that the various observational
estimates of SN-Ia rates have systematically
erred on the high side. A SN-Ia rate may be systematically high due
to contamination of a type-Ia sample by, e.g., core-collapse SNe or
active galactic nuclei. In fact, such contamination may be behind
some of the discrepant SN-Ia rates measured at $z\sim 0.7$ by several 
groups (see discussion in 
Poznanski et al. 2007, and references therein), which lead
to a fairly large range in the estimates of $\eta_B$. However, I have 
shown that even the lower estimates of $\eta$ are still high, compared
with the theoretical expectations, and furthermore there is
consistency with intermediate values of $\eta\approx 15\%$ from a variety
of independent methods (not all of them involving SN rates). SN rates can
also be overestimated if the sensitivity times of the surveys are
underestimated, i.e., the surveys are more sensitive to faint SNe than
they assume, which is quite unlikely to always be the case.
  
\subsubsection{Mass-ratio distribution}
A flat distribution in the binary mass-ratio, $f(q)$, was assumed
for close binaries in the Tutukov \& Yungelson (2002) simulations,
while in the observational analysis I considered distributions that 
are either flat or mildly rising toward low $q$. 
An inverted mass ratio distribution that favors large ($\approx 1$)
values of $q$ would
raise $f_b$, and thus lower the ``observed'' $f_{Ia}$ (in an extreme
version, $f(q)$ would be a $\delta$ function at $q=1$, i.e., all
close binaries are ``twins''). In the simulations, such 
``twinness'' would increase the rate of occurrence  of
super-Chandra mergers (although such equal-mass mergers would also be more
prone to dynamical instability leading to core collapse, rather than
a SN-Ia; Nomoto \& Iben 1985). There have been some 
indications for a peak in $f(q)$ at $q=1$ for short-period
binaries (e.g., Tokovinin
2000; Halbwachs et al. 2003; Pinsonneault \& Stanek 2006),
but it has been counter-argued these are due to 
observational selection effects, and are
not supported by other studies (see above).
Pinsonneault \& Stanek (2006) find that 55\%  of a sample of
detached binary systems that they study is in a ``twin'' peak in the
$q$ distribution. However, their twins all have primaries in the $m=12-23$ mass
range, so even if this is not a selection effect, the relevance
to the present study is not certain. 

\subsubsection{Binary separation distribution}
A similar type of solution is to assume that there are more close binaries
than indicated by modern studies, and this would again work toward
making the observational and theoretical estimates of $f_{Ia}$ 
more consistent with each other. Again, there is no solid observational
evidence, or theoretical motivation,
 for such a deviation from the fairly well-established Oepik 
distribution.

\subsubsection{A ``middle-heavy'' IMF?} 
An alternate approach to solve the problem 
may be to hypothesize a quite different form 
for the IMF. For example, a ``middle-heavy'' IMF with a ``bump'' 
at intermediate masses could provide a larger pool of SN-Ia
progenitors, and would lower the observational estimates of $f_{Ia}$.
Interestingly, Fardal et al. (2007) have recently 
proposed such a ``paunchy'' IMF, from a completely different
motivation. They found that measurements of the  extragalactic 
background light (EBL) and of the cosmic star formation history, $SFR(t)$,
may overpredict the observed local stellar density, if
one of the standard IMFs is assumed. As an example of a paunchy
IMF that can solve this problem, they proposed an IMF with a 
slope $\alpha=-1.7$ in the $m=0.5$ to $m=4$ range, falling to
$\alpha=-2.7$ at higher masses. They noted the difficulties of
measuring the IMF in this intermediate-mass region (Kroupa 2002),
and the fact that Sirianni et al. (2000) had obtained, for the R136
cluster in the Large Magellanic Cloud (LMC),
 an IMF slope of $\alpha=-1.3$ in the $1.3<m<2.1$ range.
The IMF proposed by Fardal et al. (2007), when normalized to match the
Salpeter IMF at $m=1$, has 1.5 times as many $m=3-8$ stars as the Salpeter
case, and $\eta_{\rm A}$ and $\eta_{Fe}$ would be reduced by a factor
$0.68$.  If normalized at $m=8$, the reduction, which would apply
to $\eta_B$, is only by a factor $0.96$. The ratio of core-collapse
to Ia progenitors is reduced by 0.74, compared to the Salpeter case,
and this would be the effect on $\eta_{\rm Ia/CC}$ and on $\eta_{\rm abund}$.
Thus, the effect of this particular paunchy IMF is overall moderate, and 
a more extreme variant is required 
to reduce $\eta$ more significantly. 

However, direct IMF measurements in star-forming regions generally
do not show such a feature. For example, Preibisch et al. (2002)
measured the IMF in the $m=0.1-20$ range for the Upper Scorpius OB
association.
They found the IMF to be broadly consistent, in the $m>1$
range, with the standard Scalo (1998) and Kroupa (2002) IMFs. Their
best fit slope was $\alpha=-2.8\pm 0.5$ in the $m=0.6-2$ range, and
$\alpha=-2.6\pm 0.3$ in the $m=2-20$ range. 
There have been other claims for top-heavy IMFs in the cores
of dense young clusters in our Galaxy and in the LMC, 
only to be contradicted by other studies of the same objects. For
example, Stolte
et al. (2005) found for the Arches cluster an extremely top-heavy IMF,
with $\alpha=-1.7$ for $m=6-16$, flattening to $\alpha\sim -1$ below
$m=6$. Kim et
al. (2006), in contrast, deduced for the same cluster an initial
$\alpha$ of $-2.0$ to $-2.1$ in the $m=1.3-50$ range, while  
Espinoza et al. (2007) find 
$\alpha=-2.4\pm 0.1$ in the $m=4-90$ range, fully consistent with a
Salpeter slope. Portegies Zwart et al. (2007) have recently used
N-body simulations to argue that all of the data for the Arches
cluster are consistent with a Salpeter IMF in the $m=1-100$ range,
but that the results for this and other young clusters 
are distorted by a combination of dynamical and
observational effects. Similar consistency with a Salpeter slope
has been found in other clusters (e.g., Ascenso et al. 2007).
In the Sag A* cluster in the Galactic center, Nayakshin \& Sunyaev
(2005) have used the low level of X-ray emission to deduce a
 factor-10 deficiency in the number of $m<3$ stars,
compared to the Orion Nebula cluster, and to expectations from a
standard IMF. Near-infrared observations by
 Maness et al. (2007) lead to similar conclusions.
 However, this case of top-heavy star formation
 is thought to be a peculiarity 
related to the proximity of the Sag A* cluster to 
the Galaxy's supermassive black hole. 
% No
%such X-ray deficiency is observed for the Arches cluster. 
Thus, it appears that direct and persistent evidence for a non-standard
IMF is scarce, and this is unlikely to be the solution the SN-Ia 
formation efficiency problem. 

\subsubsection{Problems with the theoretical models}
Turning to the theoretical calculations, as noted above, Tout (2005)
has made the point that values of $\alpha_{\rm CE}$ sufficiently 
larger than unity can improve dramatically the efficiency of both the 
SD and DD channels. In his simulations, 
raising $\alpha_{\rm CE}$ from 1 to 3 increases
the SN-Ia rate by an order of magnitude, in some cases. If such an
assumption is valid, this could be a solution to the problem. 

Finally, we may consider the more radical 
possibility that some 
of the more basic assumptions about SN-Ia progenitors are wrong.
Both the DD and SD scenarios are not free of theoretical and
observational problems. It has long been argued that the final stage
of a DD merger will lead to an accretion-induced core collapse, rather
than a SN-Ia (e.g., Nomoto \& Iben 1985; 
Saio \& Nomoto 2004, Guerrero et al. 2004), 
although there are 
opposing views that invoke rotation of the stellar surface to prevent
this outcome (e.g., Piersanti et al. 2003).
An observational
search for DD progenitor systems (Napiwotzki et al. 2004;
Nelemans et al. 2005) has turned up, among $\sim 1000$ WDs surveyed, 
few or no potential DD systems
with a total mass exceeding $M_{\rm ch}$ that will merge within a Hubble
time. It is not yet clear if this low detection rate is significantly
smaller than expected if the DD scenario is to explain  
the Galactic SN-Ia rate (which is, itself, quite uncertain).
For example, if the Galactic rate is $10^{-3} {\rm yr}^{-1}$, there
should be $10^7$ progenitor systems that will merge within
$10^{10}$~yr. If there are $10^{11}$ stars in the Galaxy, then, assuming
a Kroupa (2001) IMF, of order 10\%, or $10^{10}$,  will be WDs.
Only one in a thousand WDs surveyed would therefore be expected to be a SN-Ia
progenitor, which is consistent, at this order-of-magnitude level,
with the current survey. A more detailed calculation, 
taking into account Galactic structure and star formation
history, survey completeness, and selection effects, is required in
order to determine whether or not the observations are in conflict
with the DD scenario.  

The SD scenario, in turn, has been criticized 
for its assumptions about the existence of {\it ad hoc} mechanisms that
regulate the accretion flow on to the primary (e.g., Cassisi et
al. 1998; Piersanti et al. 1999, 2000). Observationally,
Badenes et al. (2007) have noted the absence, 
in seven nearby SN-Ia remnants, of the
signatures of the strong wind from the accretor that supposedly
stabilises the accretion flow, 
and permits reaching $M_{\rm ch}$ in the Hachisu et al. (1996) model. 
Prieto et al. (2007) have pointed out that a subtantial number of
SNe-Ia have been discovered, by now, in low-metallicity galaxies, in
contrast to to the prediction by Kobayashi et al. (1998) that 
SNe-Ia cannot explode in such environments, due to a minimum
metallicity that is required for the wind regulation mechanism 
to be effective. (Yoon et al. 2004 have proposed WD rotation as an 
alternative or additional mechanism for stabilising high accretion
rates.) While evidence for circumstellar material, consistent with
expectations from a wind from a red-giant companion, has been found in
one recent normal SN-Ia (Patat et al. 2007), such material is not
observed in another event (Simon et al. 2007). 
For the remnant of Tycho's SN, which 
was a type-Ia (Badenes et al. 2006), there have been
conflicting claims about the identification and nature of a remaining companion
star (Ruiz-Lapuente et al. 2004;  Fuhrmann 2005; Ihara et al. 2007).
If, in the end, no companion is found, this would be another problem
for the SD picture. 
 
\subsubsection{Single-star SN-Ia progenitors?} 
A speculative direction that may be worth considering is that, 
perhaps, SNe-Ia are not descended from binary systems.
Iben \& Renzini (1983)
have reviewed several stellar evolution studies
of intermediate-mass stars, in which mass loss has not reduced the
remaining stellar mass below $\sim M_{\rm ch}$ at the time of carbon ignition 
in the core. Such single stars may undergo a thermonuclear
runaway that disrupts the star, but Iben \& Renzini (1983) concluded
that, due to the many assumptions and approximations, ``one is not
left with an overwhelming sense of confidence in the detailed results''. 
Such an explosion would be expected to take place still within a
hydrogen-rich envelope, and hence would not resemble a normal SN-Ia.  
However, Tout (2005) has sketched the possibility that
 some not-fully understood aspect of stellar evolution 
leads all stars in some initial mass range to avoid mass loss
on the red giant branch, but to then completely lose
their hydrogen envelopes on the asymptotic giant branch (AGB), 
proceeding then to a Chandra-mass thermonuclear 
runaway of their cores in such a ``single-star'' SN-Ia channel.
In a variant of this idea, Waldman, Yungelson, \& Barkat (2007) have
postulated that interaction with a binary companion is responsible for
the stripping of the envelope, but the stripped star then goes on to 
explode as a single star. 

Observationally, recent metal 
abundance measurements in Kepler's SN remnant strongly suggest 
a type-Ia explosion (e.g., Blair et al. 2007; Reynolds et al. 2007). 
However, the $\sim 300$~km~s$^{-1}$ 
velocity of the remnant as a whole away from the Galactic plane,
together with evidence for a circumstellar medium from a massive
progenitor, are difficult to understand in the context of a binary progenitor
system (Reynolds et al. 2007). Aubourg et al. (2007) have modeled the 
stellar populations of SN-Ia host galaxies, and concluded that the prompt 
SN-Ia component explodes within 70~Myr of star formation, implying a
primary initial mass range of $m=6-8$.\footnote{The SN sample of
  Aubourg et al. (2007) is a literature compilation from several SN
  surveys and from IAU circulars, 
for which selection effects and ``control times'' (the time
during which a SN-Ia could have been detected in a given galaxy) are
not known. In their derivation of a SN-Ia rate as a function of 
star formation rate, these authors implicitly assume that the control
times and the followup efficiencies 
are the same for SNe in both the star-forming and the quiescent galaxies in the
surveys that produced the SN sample. At some level, this assumption 
likely fails, and this casts some doubt on the accuracy of their
result. For example, star-forming galaxies tend to host more luminous,
more slowly fading, SNe-Ia, compared to early-type galaxies
(see, e.g., Gallagher et al. 2005); such SNe
are more likely to be detected, and more likely to be confirmed 
spectroscopically as type-Ia's, artificially biasing high the derived SN-Ia
rate in star-forming galaxies. If some of the surveys specifically
targeted late-type galaxies, the bias would be even greater.}
For such a narrow and high range, all of the 
estimates of $\eta$ through the prompt channel will become
even higher than found earlier. For example, $\eta_{Ia/CC}$ will be
$67\pm 15\%$. Even with the most conservative binary parameters
(including a substantial ``twin'' population, which could raise $f_b$,
but at the same time would restore the duplicity factor, $f_d=1/2$), 
it is then difficult to
avoid an excess of SNe-Ia compared to progenitor binary systems.
A single-star channel would solve this problem.
 
A good place to test for the existence of
 such single-star progenitors could be the
LMC. Taking a value of $B\sim 1\times 10^{-3} M_{\odot}^{-1}$ for the SN-Ia 
rate per star formation rate (see \S2.2), which should be appropriate for
star-forming dwarf galaxies like the LMC,  
and a $SFR$ value of $0.3 M_{\odot} {\rm yr}^{-1}$ (Kennicutt
et al. 1995), an LMC SN-Ia rate of $\sim 3\times 10^{-4}
{\rm yr}^{-1}$ is obtained. Another estimate can be made if 
the 2:1 to 4:1 ratio (Mannucci et al. 2005) of core-collapse to SN-Ia rates
in star-forming galaxies holds for the LMC, then amongst the 45 to 76 SN
 remnants in the LMC (Sasaki, Haberl, \& Pietsch 2000; Bojicic et
 al. 2007), roughly $9-25$ should be the remnants of SNe-Ia. 
Divided by a typical $10^4$~yr lifetime of a remnant,
this gives a SN-Ia rate of $\sim (9-25)\times 10^{-4}
{\rm yr}^{-1}$. The AGB stage of ZAMS $m=6-7$ stars 
lasts of order $10^5$~yr (e.g., Girardi et al. 2000). 
Thus, if SN-Ia progenitors are AGB stars that have lost their
 hydrogen  envelopes, there could be a few tens or hundreds of 
such stars in the LMC. 
Calculated colors and spectra of such objects could be
compared to observations by searching for candidate progenitors
in photometric 
catalogues of luminous stars in the LMC
(Cioni et al. 2000; Nikolaev \& Weinberg 2000).  

To summarize, I have collected measurements pertinent to five
independent methods for estimating the fraction of potential SN-Ia
progenitors that explode as SNe-Ia. I have compared them
self-consistently, and studied the dependence of the fraction on 
the assumed IMF and on other parameters.  
Modern measurements of SN rates and related observables 
leave little doubt that 
a large fraction of the intermediate-mass stellar population
explodes as SNe-Ia, larger than would be expected based on 
most SD and DD models. 
A critical reappraisal of our ideas about SN-Ia progenitors may be in order.
   
\section*{Acknowledgments}
I thank Lev Yungelson for providing some unpublished statistics
from his simulations, and Carles Badenes, 
Wolfgang Hillebrandt, Mordecai Mac Low, 
Tsevi Mazeh, Paolo Mazzali, Jacco Vink, and Dennis Zaritsky
for comments and advice.
The Kavli Institute for Theoretical Physics, where this
work was initiated, is thanked for its hospitality.
This research was supported in part by the National Science
Foundation under Grant No. PHY05-51164.

%\onecolumn
%\newpage

\end{document}